\documentclass[onecolumn,showpacs,preprintnumbers,amsmath,amssymb]{revtex4}

\usepackage{lipsum}
\usepackage{tabularx}
\usepackage{array}
\usepackage{mathrsfs}
\usepackage{amssymb}
\usepackage{amsmath}
\usepackage{cases}  
\usepackage{graphicx}
\usepackage{subfigure} 
\usepackage{dcolumn}
\usepackage{bm}
\usepackage{verbatim} 
\usepackage[dvipdfm,pdfstartview=FitH,colorlinks,linkcolor=blue,anchorcolor=blue,citecolor=blue]{hyperref}
\usepackage{amsthm}  

\begin{document}

\title{Non-Markov property of afterpulsing effect in single-photon avalanche detector}

\author{Fang-Xiang Wang, Wei Chen,\footnote{weich@ustc.edu.cn} Ya-Ping Li, De-Yong He, Chao Wang, Yun-Guang Han, Shuang Wang, Zhen-Qiang Yin and Zheng-Fu Han}
\address{Key Laboratory of Quantum Information, University of Science and Technology of China, Hefei 230026, China\\
and Synergetic Innovation Center of Quantum Information $\&$ Quantum Physics, University of Science and Technology of China,\\
Hefei, Anhui 230026, China\\}


\begin{abstract}

Single-photon avalanche photodiode(SPAD) has been widely used in researching of quantum optics. Afterpulsing effect, which is an intrinsic character of SPAD, affects the system performance in most of the experiments and needs to be carefully handled. For a long time, afterpulsing has been presumed to be determined by the pre-ignition avalanche. We studied the afterpulsing effect of a commercial InGaAs/InP SPAD (APD: Princeton Lightwave PGA-300) and demonstrated that its afterpulsing is non-Markov, which has memory effect of the avalanching history. Theoretical analysis and the experimental results clearly indicate that the embodiment of this memory effect is the afterpulsing probability, which increases as the number of ignition-avalanche pulses increase. The conclusion makes the principle of afterpulsing effect clearer and is instructive to the manufacturing processes and afterpulsing evaluation of high-count-rate SPADs. It can also be regarded as an fundamental premise to handle the afterpulsing signals in many applications, such as quantum communication and quantum random number generator.

\end{abstract}



\maketitle

\section{Introduction}
\label{intro.}
Avalanche photodiode (APD) is widely used in broad fields, such as single molecule detection \cite{Li1993,Moerner2003}, autocorrelated fluorescent decays and quantum information \cite{Eisaman2011,Hadfield2009}, to detect extremely weak light signals or even single photons. The latter is usually named single-photon avalanche detector (SPAD), which acts as an essential device and may greatly influence the performance of the system in many applications, especially in quantum information fields \cite{Hadfield2009,Comandar2014}. SPADs usually works in two different modes:free-running mode and gated mode. For free-running mode, the SPAD is biased above its breakdown voltage to allow an avalanche happens. For gated mode, the SPAD is commonly biased below its breakdown voltage and avalanche process can only happen in gating time windows when additional gate voltage signals are superimposed. There are many parameters can be used to quantify the performance of SPADs, such as clocking rate, dark count rate, detection efficiency, timing jitter and so on. In the past decade, the performance of SPAD has been greatly improved and can working at gigahertz clocking frequency with up to one hundred megahertz counting rates\cite{Yuan2007,Yuan2012,Scarcella2013}.

However, afterpulsing effect of SPAD severely limits the count rate and applications in precise measurements. Afterpulsing, which is correlated with the ignition avalanche, comes from detrapping of carriers that were trapped by deep energy level in the junction depletion layer \cite{Haitz1965,Cova1991,Cova1996,Kang2003}. The trapped carriers then release from the traps over time by thermal fluctuation. Usually, the release lifetime $\tau$ is dependent on the type of traps and can vary from 10 $ns$ to several microseconds \cite{Cova1991,Cova2003,Jensen2006}, which is usually much longer than that needed for quenching an avalanche. A releasing carrier may initiate another pseudo detection signal, which is usually called afterpulsing or after pulse, if the SPAD is biased above the breakdown voltage. The pseudo signal leads to negative effects on SPADs' applications. For example, afterpulsing will increase quantum bit error rate in quantum key distribution \cite{Yoshizawa2003,Jain2015} and destruct the randomness of quantum random number generator\cite{Wang2015}. In order to reduce afterpulsing, another hold-off time depending on $\tau$ is necessary. The afterpulsing effect is negligible if the hold-off time is much longer than carriers' lifetime $\tau$. But it is not practical for high-count-rate SPADs. Many approaches have been proposed to reduce afterpulsing \cite{Hadfield2009,Cova1996,Yuan2007,Tosi2014,Wayne2014,Scarcella2015}, but it remains a critical limitation for high-count-rate SPADs. Precisely modeling the behavior of afterpulsing is the foundation to handle this problem.

\section{Theoretical analysis of afterpulsing effect}
\label{theory}

The trap lifetime $\tau$ of APD can be described by the Arrhenius equation 
\begin{equation}
\tau= C e^{E_{a}/{kT}}
\label{equ1}
\end{equation}
where $T$ is the temperature, $k$ is the Boltzmann constant and $C$ is determined by the relevant effective states density, cross section of the trap and temperature $T$ \cite{Jensen2006}. The carriers releasing probability is then exponential via time $t$. As the types of traps in APD are dependent on practical manufacturing processes, there are thereby variety of traps with different lifetime $\tau_j$. Thus, the afterpulsing probability (AP) during a detection window $\Delta t$ contains different exponential components \cite{Cova1991}
\begin{equation}
p_a(j)=A_{1}exp(-\frac{j\cdot \Delta t}{\tau_1})+A_{2}exp(-\frac{j\cdot \Delta t}{\tau_2})+\cdots.
\label{equ2}
\end{equation}
where $A_i$ are amplitudes of different exponential components. and $p_a(j)$ means the AP of the $j$-th detection window after the ignition avalanche. Recently, different fitting methods have been studied, which employed a broad distribution of different exponential components instead of summating a few dominate discrete ones \cite{Itzler2012,Horoshko2014}.

So far, most studies on afterpulsing effect just focus on its time-dependent probability distribution ignited by a single light pulse, and do not discriminate the situation with single ignition avalanche from that with multiple ones \cite{Yoshizawa2002,Kang2003,Yen2008,Silva2011,Stipcevic2013}. However, it is not the real situation when the SPAD works continuously for photon detection. In this case, several avalanches may happen successively within the lifetimes of trapped carriers. Previous works have demonstrated that traps filled with carriers during an avalanche event are much smaller than the total number of traps \cite{Cova1991,Jensen2006,Liu1992}. Thus, avalanches happen successively will increase the number of filled traps and hence will enhance the AP. That is, the AP distribution may be correlated to the avalanching history, while prior studies have not delve down into this physical procedure.

In this paper, we studied the afterpulsing influence of multi-ignition avalanches on photon detection process of InGaAs/InP SPAD under different gate frequencies and photon intensities and found out that afterpulsing possesses memory effect and thus correlates to the avalanching history. We call this property as non-Markov property \cite{Bharucha-Reid2012,Markovprocess}. We gave out analytical expressions of afterpulsing influences on photon detection in theory. The afterpulsing influence on photon detection can then be quantified. We also demonstrated experiments to verify the theory on afterpulsing effect. The experimental results fitted well with the theoretical model. The model is instructive to the manufacturing processes and afterpulsing evaluation of high-count-rate SPADs. It also offers a new perspective to deal with practical security of quantum communication with imperfect devices \cite{Gottesman2004,Renner2008,Scarani2009}.

\subsection{Markovian Model}
\label{Markov-like}

Considering that trap level distribution is a technological parameter, which is determined once the product is fabricated and the details have been studied well, we do not deal with specific time-dependent distribution of AP in the next analysis process. The primary data of $p_{a}$ can be obtained by experimental measurements. In addition, It is reasonable to assume that the filling and release processes of different traps are independent as the filling is rare.

We model the procedure according to Markovian and Non-Markov procedure individually. Firstly, we assume that AP distribution is only determined by the latest ignition avalanche as assumed in \cite{Humer2015},namely, the AP distribution has no memory of prior avalanching history and is Markovian. The probability of an avalanche triggered by laser pulse and dark count can be derived from the Poisson distribution and avalanche probabilities of different gates (detection windows) are independent identical distribution (IID) \cite{Wang2015}. Let $p$ be the avalanche probability per gate triggered by photon pulse, $p_a(j)$ be the AP of Gate-$j$ with single-ignition photon pulse and $p_n$ be the total avalanche probability of Gate-$n$. There is no afterpulsing for the first gate. As we assume that afterpulsing effect comes from the latest ignition avalanche only, the system should have no memory of avalanching history before that.
That is, if Gate-$(n-1)$ avalanches, $p(n|n-1)=p+(1-p)p_a(1)$. If Gate-$(n-1)$ does not avalanche, Gate-$(n-2)$ is considered. If Gate-$(n-2)$ avalanches, $(n|(\overline{n-1},n-2))=p+(1-p)p_a(2)$. If Gate-$(n-2)$ does not avalanche, Gate-$(n-3)$ is considered, and so on. Hence, the average avalanche probability of Gate-$n$ is
\begin{equation}
\begin{aligned}
p_n=&\quad p_{n-1}[p+(1-p)p_a(1)]\\
&+(1-p_{n-1})\{p_{n-2}[p+(1-p)p_a(2)]+(1-p_{n-2})\\
&\quad\{\cdots\{p_0[p+(1-p)p_a(n)]+(1-p_0)p\underbrace{\}\cdots\}\}}_{(n-1)"\}"}.
\end{aligned}
\label{equ3}
\end{equation}
The first term of the equation means if the pre-gate avalanches, the ones before the pre-one are not considered. If the pre-gate does not avalanche,  the gate before the pre-gate is under considered, and so on. If AP $p_a(j)$ is small, and only the first order terms are reserved, namely, $p_a(i)p_a(j)\cong 0$, then the approximate equation becomes
\begin{equation}
p_n\cong p[1+\sum\limits_{j=1}^{n} (1-p)^{j}p_a(j)].
\label{equ4}
\end{equation}
See appendix for detailed proof of Equation \ref{equ4}.

\subsection{Non-Markov Model}
\label{non-Markov}
Then, we assume that the system has memory of prior avalanching history and AP distribution is determined by the avalanching history. Thus, the derivation of avalanche probability of Gate-$n$ is much different from that of Markov-like Model and the full-expression depending on the avalanching history is very complex. For example, $p_1=p[p+(1-p)p_a(1)]+(1-p)p=p[1+(1-p)p_a(1)]$, $p_2=p[p+(1-p)p_a(1)]\{p+(1-p)[p_a(1)+p_a(2)]\}+p\{1-[p+(1-p)p_a(1)]\}[p+(1-p)p_a(2)]+(1-p)p_1$. Here we give a formal expression of $p_n$, and then give the approximate equation of $p_n$. $p_n$ can be classified into two sets: Gate-0 avalanches in the first set $F(p,p_a(j))$ and does not avalanche in the second set $S(p,p_a(j))$. The second set can be expressed as $S(p,p_a(j))=(1-p)p_{n-1}$. Then $p_n$ can be expressed as
\begin{equation}
p_n= F(p,p_a(j))+S(p,p_a(j))
\label{equ5}
\end{equation}

By adopting the same approximation utilized in the Markov model, where $p_a(i)p_a(j)\cong 0$, then we obtain the approximate equation of Equation \ref{equ5},
\begin{equation}
p_n\cong p[1+\sum\limits_{j=1}^{n}(1-p)p_a(j)].
\label{equ6}
\end{equation}
See appendix for detailed proof of Equation \ref{equ6}. In fact, the formal expression is enough for the proof.

Comparing Equation \ref{equ6} with Equation \ref{equ4}, we find that, for both cases, $p_n$ depends on two parameters. The first one is $p$, the photon-ignition avalanche probability without afterpulsing. The second one is the AP distribution $p_a(j)$. The afterpulsing influence increases via $n$, but the weight of $p_a(j)$ decays exponentially. The decay rate depends on the base $(1-p)$. In fact, the afterpulsing influence becomes saturation when $n$ is large enough. It is reasonable to truncate $n$ to an appropriate value according to the actual value of $p$. The equation will degenerate into $p_n=p$ if no afterpulsing exists.  However, the influence of afterpulsing of non-Markov model is more significant. The weight of $p_a(j)$ in non-Markov case is $(1-p) $, which is linear. But the corresponding weight of $p_a(j)$ in Markov-like case is $(1-p)^j$, which decays exponentially. Thus, the Markovian approximation holds only for $p,p_a(j)\ll 1$. The difference between these two model will become more and more significant as $p$ or  $p_a(j)$ (or both of them) increases. The experiments demonstrated below prove that afterpulsing effect is non-Markov. 

\section{Experiments and discussion}
\begin{figure}[t]
\centering
\resizebox{8cm}{4.5cm}{\includegraphics{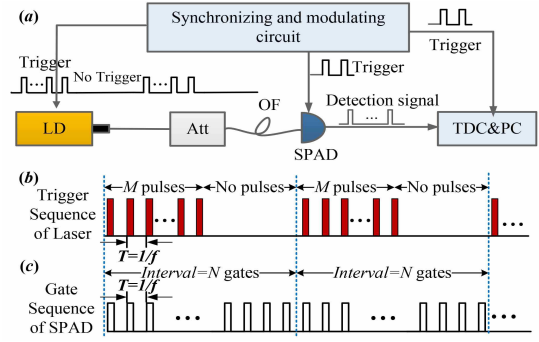}}
\caption{(Color online) The schematic measurement setup of non-Markov effect. LD: laser diode; Att: adjustable attenuator; OF: optical fiber; TDC: time-to-digital converter; PC: personal computer.}
\label{fig:setup}
\end{figure}

The experimental setup is shown in Figure \ref{fig:setup}. The laser diode (LD) is operated in pulsed mode and is triggered by a modulating signal. Figure \ref{fig:setup}(b) shows the modulating signal, which includes $M$ trigger pulses for every interval $\Delta T$. The trigger frequency $f$ and interval time $\Delta T$ are adjustable. The SPAD (The model of APD is: Princeton Lightwave PGA-300) is operated in gated mode without dead time and the gate width is 2.5 $ns$. The detection efficiency $\eta$ of the SPAD is about 10.5\%. The detection gate is triggered by external pulsed signal, as shown in Figure \ref{fig:setup}(c). All trigger signals are synchronized. Laser pulses from the LD are attenuated by a adjustable attenuator and then sent into the SPAD. The SPAD outputs a detection signal if an avalanche happens. The detection signals are input into a time-to-digital converter (TDC, Agilent U1051A Acqiris TC890) in real time. The TDC can records the arriving time of every detection signal with a resolution of 50 $ps$. The parameter $p$ can be modulated by adjusting the attenuator. The parameter $p_a(j)$ is modulated by changing the trigger frequency $f$.

The verification work consists of three  experiments. \textbf{In the first experiment}, the AP distributions under different trigger frequencies ($f=2, 5, 10$ and $20$ MHz) are measured with single ($M=1$) and multi-ignition laser pulses ($M\geqslant 100$). The average photon number $\lambda$ per pulse is set as 0.07. The AP $p_a(j)$ is the avalanche probability of Gate-$(M+j-1)$ after subtracting the dark count. \textbf{In the second experiment}, the average photon number $\lambda$ is modulated ($\lambda=1.0, 0.7, 0.07$ and $0.02$) so that the avalanche probability $p_n$ with different $p$ can be measured, while the AP distribution $p_a(j)$ varies little. \textbf{In the third experiment}, the afterpulsing distribution $p_a(j)$ is modulated. The trigger frequency $f$ is modulated ($f=2, 5, 10$ and $20$ MHz) so that the AP distribution $p_a(j)$ becomes different, while $\lambda$ is set as constant ($\lambda=0.07$). For all experiments, the dark count rate per gate is under the order of $10^{-5}$.

\begin{figure}[t]
\centering
\resizebox{8.5cm}{4.5cm}{\includegraphics{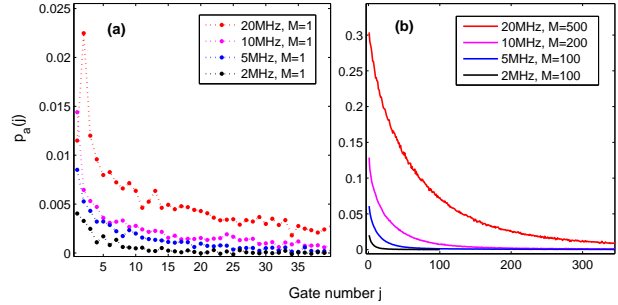}}
\caption{(Color online) The afterpulsing probability distributions with (a) single and (b) multi-ignition pulses. The average photon number per laser pulse is $\lambda=0.07$.}
\label{fig:afpprobdist}
\end{figure}

The first experiment demonstrates that AP increases as $f$ increases (see Figure \ref{fig:afpprobdist}), where $\lambda=0.07$. As shown in Figure \ref{fig:afpprobdist}(a), $p_a=2.0\%$ when $f=2$ MHz, while $p_a=29.9\%$ when $f=20$ MHz, where $p_a(j)$ is the AP of Gate-$j$ and $p_a$ is the total AP with single-ignition pulse. It also demonstrates that APs with multi-ignition pulses (Figure \ref{fig:afpprobdist}(b)) are much larger than that with single ignition pulse (Figure \ref{fig:afpprobdist}(a)). For example, when $f=5$ MHz, $p_{a,100}(1)=6.1\%$, while $p_a(1)=0.9\%$ and $p_a\approx\sum\limits_{j=1}^{M=100}p_a(j)=6.2\%$, where $p_{a,M}(j)$ is the AP of Gate-$(M+j-1)$ with $M$-ignition pulses. Thus, $p_{a,100}(1)\approx p_a$. It means that AP accumulates as $M$ increases. Namely, AP depends on the avalanching history and is non-Markovian. The AP becomes larger as $f$ increases. As the full width at half maximum (FWHM) of the detection signal output from the SPAD is about 40 $ns$, the saturated processing rate of the discrimination circuit is less than 20 MHz by taking the processing time into account. Thus, $p_a(1)$ drops and is even smaller than expected when $f=20$ MHz. However, the processing rate limitation is not that significant for the experiments because the average photon number $\lambda=0.07$, which is small, when $f=20$ MHz.

\begin{figure}[t]
\centering
\resizebox{8.5cm}{5.5cm}{\includegraphics{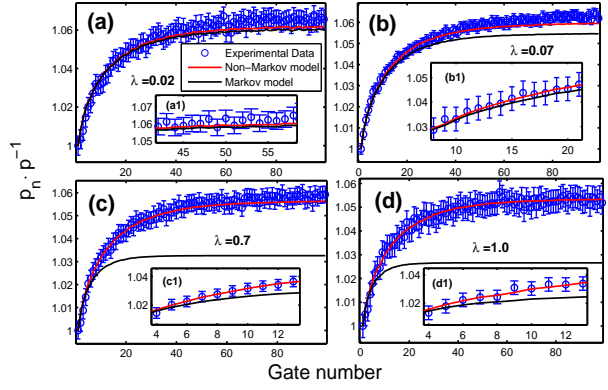}}
\caption{(Color online) The experimental data and theoretical simulations for $p_n$ with different average photon number. The trigger frequency is $f=5$ MHz. The insets are the partial enlargements. The error bar is set as $3\sigma$, where $\sigma$ is the standard deviation of statistics.}
\label{fig:different pa}
\end{figure}

The second experiment (see Figure \ref{fig:different pa}) shows that the difference between non-Markov model and Markov model is negligible when $\lambda=0.02$ ($p=1-e^{-\eta\lambda}\approx 0.002$). The difference is significant when $p$ is larger, e. g., Figure \ref{fig:different pa}(d) shows that $p_{{}_{n=99}}^{non-Markov}-p_{{}_{n=99}}^{Markov}=1.053\cdot p-1.027\cdot p=0.026\times p$. The Markov model can no longer fit with the experimental data. 

\begin{figure}[t]
\centering
\resizebox{8.5cm}{5.5cm}{\includegraphics{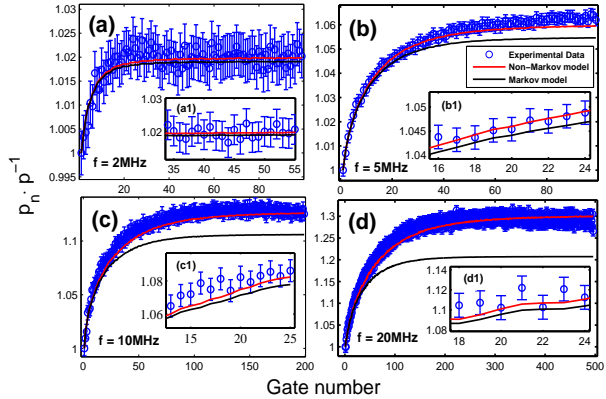}}
\caption{(Color online) The experimental data and theoretical simulations for $p_n$ with different gate frequency $f$. The average photon number per laser pulse is $\lambda=0.07$. The detection efficiencies are $\eta_{{}_{f=2MHz}}=10.3\%,\eta_{{}_{f=5MHz}}=10.5\%,\eta_{{}_{f=10MHz}}=11.6\%,\eta_{{}_{f=20MHz}}=12.0\%$. The insets are the partial enlargements. The error bar is set as $3\sigma$, where $\sigma$ is the standard deviation of statistics.}
\label{fig:different gate}
\end{figure}

The third experiment gives the differences between non-Markov model and Markov model with different $p_a(j)$ (see Figure \ref{fig:different gate}). There is hardly any difference between these two model when $f=2$ MHz ($p_a=2.0\%$), as shown in Figure \ref{fig:different gate}(a). But the difference becomes $p_{n}^{non-Markov}-p_{n}^{Markov}\approx 0.1\cdot p$ for large $n$ when $f=20$ MHz ($p_a=29.9\%$).

The experimental results above are all consistent well with the non-Markov model. According to the Equation \ref{equ6}, the afterpulsing influence on photon detection depends on the product of $p$, the pulse-ignition avalanche probability per detection window, and $p_a$, the total AP. In Figures \ref{fig:different pa}(a), $p\cdot p_a=1.3\times 10^{-4}$. And in Figure \ref{fig:different gate}(a), $p\cdot p_a=1.4\times 10^{-4}$. In both figures, the Markov model also fits well with the experimental data. But in Figures \ref{fig:different pa}(b),(c),(d) and \ref{fig:different gate}(b),(c),(d), Markov model can no longer fit with the data, where $p\cdot p_a$ is $5.3\times 10^{-3}$ in Figures \ref{fig:different pa}(d) and is $2.1\times 10^{-3}$ in Figures \ref{fig:different gate}(d). The non-Markov effect will be more significant for higher $p_a$. It is really the case for high-count-rate SPADs, of which the detection period is much smaller than the lifetime of carriers \cite{Yuan2007,Yuan2012,Scarcella2013}.

It should be note that detection efficiency of the SPAD varies when trigger frequency $f$ varies. In order to make comparing of AP influence with different $f$, the gate voltage loaded on SPAD is fine adjusted so that detection efficiency varies less ($\eta_{{}_{f=2MHz}}=10.3\%,\eta_{{}_{f=5MHz}}=10.5\%,\eta_{{}_{f=10MHz}}=11.6\%,\eta_{{}_{f=20MHz}}=12.0\%$). Otherwise, afterpulsing would be a catastrophe and the SPAD under $f=20$ MHz could no longer work properly if the voltage remains the same with that under $f=2$ MHz.

\section{Conclusion}
\label{conclusion}

In conclusion, we have proved that afterpulsing effect of SPAD has memory on avalanching history and is non-Markov. We give an approximate expression (Equation \ref{equ6}) that describes the non-Markov effect well. The theoretical analysis fits well with the experimental results. The non-Markov effect will be more significant for higher $p_a$. It is really the case for high-count-rate SPADs, of which the detection period is much smaller than the lifetime of carriers \cite{Yuan2007,Yuan2012,Scarcella2013}. As it is the embodiment of rarely filling of traps, the non-Markov effect of afterpulsing should be general to different types of SPADs provided that the filling is rare during an avalanching process. Our work makes the principle of afterpulsing effect clearer and is instructive to the manufacturing processes and afterpulsing evaluation of high-count-rate SPADs. It also offers a new perspective to deal with practical security of quantum communication due to the imperfection of devices.

\section*{Acknowledgments}
\label{Acknowledgments}

This work has been supported by the National Basic Research Program of China (Grant Nos. 2011CBA00200 and 2011CB921200), the National Natural Science Foundation of China (Grant Nos. 61475148, 61201239, 61205118 and 11304397) and the Strategic Priority Research Program (B) of the Chinese Academy of Sciences (Grant Nos. XDB01030100 and XDB01030300).

\setcounter{equation}{0}

\appendix
\section*{Appendix}
\label{Appendix}

\subsection{The Markovian approximate equation}

The complete equation of avalanche probability per gate in Markov model is
\begin{equation}
p_n=p_{n-1}[p+(1-p)p_a(1)]+(1-p_{n-1})\{p_{n-2}[p+(1-p)p_a(2)]+(1-p_{n-2})\{\cdots\{p_0[p+(1-p)p_a(n)]+(1-p_0)p\underbrace{\}\cdots\}\}}_{(n-1)"\}"}.
\label{equ11}
\end{equation}
By keeping the first-order terms of $p_a(j)$ only, we obtain the approximate equation
\begin{equation}
p_n\cong p[1+\sum\limits_{j=1}^{n} (1-p)^{j}p_a(j)].
\label{equ12}
\end{equation}
\begin{proof}
First, we consider the limiting case of no afterpulsing. In this case, $p_n=p$ always holds.

Thus, in general, $p_n$ can be expressed as $p_n=p+g_n(p_a(j))$, where $g_n(p_a(j))=g_n(p_a(1),p_a(2),p_a(3),\cdots)$ is a power series of $p_a(j)$ and $j=1,2,\cdots$. For the first term of right side, we have
\begin{displaymath}
\begin{aligned}
p_{n-1}[p+(1-p)p_a(1)]&=[p+g_{n-1}(p_a(j))]\cdot[p+(1-p)p_a(1)]\\
& =p^2+p(1-p)p_a(1)+pg_{n-1}(p_a(j))+g_{n-1}(p_a(j))\cdot (1-p)p_a(1)\\ 
& \cong p^2+p(1-p)p_a(1)+pg_{n-1}(p_a(j)).
\end{aligned}
\end{displaymath}
The remainder terms become
\begin{displaymath}
\begin{aligned}
&(1-p_{n-1})\{p_{n-2}[p+(1-p)p_a(2)]+(1-p_{n-2})\{\cdots\{p_0[p+(1-p)p_a(n)]+(1-p_0)p\underbrace{\}\cdots\}\}}_{(n-1)"\}"}\\
=& [1-p-g_n(p_a(j))]\{p_{n-2}[p+(1-p)p_a(2)]+(1-p_{n-2})\{\cdots\{p_0[p+(1-p)p_a(n)]+(1-p_0)p\underbrace{\}\cdots\}\}}_{(n-1)"\}"}\\
=& (1-p)\{p_{n-2}[p+(1-p)p_a(2)]+(1-p_{n-2})\{\cdots\{p_0[p+(1-p)p_a(n)]+(1-p_0)p\underbrace{\}\cdots\}\}}_{(n-1)"\}"}\\
&-g_n(p_a(j))\{p_{n-2}[p+(1-p)p_a(2)]+(1-p_{n-2})\{\cdots\{p_0[p+(1-p)p_a(n)]+(1-p_0)p\underbrace{\}\cdots\}\}}_{(n-1)"\}"}.
\end{aligned}
\end{displaymath}
In order to simplify term $g_n(p_a(j))\{p_{n-2}[p+(1-p)p_a(2)]+(1-p_{n-2})\{\cdots\{p_0[p+(1-p)p_a(n)]+(1-p_0)p\underbrace{\}\cdots\}\}}_{(n-1)"\}"}$, we begin with the innermost brace. The term containing $p_a(n)$ can be abandoned directly by the approximation convention $p_a(i)p_a(j)\cong 0$. Then, the expression of the innermost brace is simplified to $p$. Then, the expression outside the innermost brace becomes $p_1[p+(1-p)p_a(n-1)]+(1-p_1)p\cong p$. The expressions of outer braces can be simplified by analogy, and $g_n(p_a(j))\{p_{n-2}[p+(1-p)p_a(2)]+(1-p_{n-2})\{\cdots\{p_0[p+(1-p)p_a(n)]+(1-p_0)p\underbrace{\}\cdots\}\}}_{(n-1)"\}"}\cong g_n(p_a(j))p$. Equation \ref{equ11} hence becomes
\begin{displaymath}
\begin{aligned}
p_n=p^2+p(1-p)p_a(1)+(1-p)\{p_{n-2}[p+(1-p)p_a(2)]+(1-p_{n-2})\{\cdots\{p_0[p+(1-p)p_a(n)]+(1-p_0)p\underbrace{\}\cdots\}\}}_{(n-1)"\}"}\\
\end{aligned}
\end{displaymath}
Analogously, $p_{n-2}, p_{n-3}, \cdots$ can be simplified by the same procedure done for $p_{n-1}$. Equation \ref{equ11} finally becomes
\begin{displaymath}
\begin{aligned}
p_n=& p^2+(1-p)p^2+(1-p)^{2}p^2+\cdots +(1-p)^{n-2}p^2+(1-p)^{n-1}p\\
&+p(1-p)p_a(1)+p(1-p)^{2}p_a(2)+\cdots+p(1-p)^{n-1}p_a(n-1)+p_0(1-p)^{n}p_a(n)\\
=& p[1+\sum\limits_{j=1}^{n}(1-p)^{j}p_a(j)].
\end{aligned}
\end{displaymath}
This completes the proof.
\end{proof}

There is an alternative proof of the Markovian approximate equation. The proof is shown below.

\begin{proof}
 The recurrence relation of the sequence $p_n$ can be written as follows according to  Equation \ref{equ11}:

\begin{equation}
p_{n+1}=p+p_{n}(1-p)p_a(1)+p_{n-1}(1-p_{n})(1-p)p_a(2)+\cdots +p(1-p_1)\cdots (1-p_{n})(1-p)p_a(n+1)
\label{equ15}
\end{equation}

According to Equation \ref{equ15}, it is easy to get that $p_1=p +p(1-p)p_a(1)$, $p_2=p+p1(1-p)p_a(1)+p(1-p1)(1-p)p_a(2)\cong p[1+(1-p)p_a(1)+(1-p)^2p_a(2)]$, where we only keep the first-order terms of $p_a(j)$.  Thus, we assume that $p_n\cong p[1+\sum\limits_{j=1}^{n} (1-p)^{j}p_a(j)]$. If $p_{n+1}=p_{n}+p(1-p)^{n+1}p_a(n+1)$ holds, then Equation \ref{equ12} holds.

According to Equation the expression of $p_{n}$, we can replace all the $p_{n}$ in Equation \ref{equ15} by $p$ since we neglect higher than the first-order terms of $p_a(j)$. Thus we can get straightforwardly:

\begin{equation}
p_{n+1}\cong p+p(1-p)p_a(1)+p(1-p)^2 p_a(2)+\cdots +p(1-p)^{n+1}p_a(n+1)=p[1+\sum\limits_{j=1}^{n+1} (1-p)^{j}p_a(j)]
\label{equ14}
\end{equation}

This means Equation \ref{equ12} holds for $n+1$, thus completes the proof.

\end{proof}

\subsection{The non-Markov approximate equation}

The formal expression of avalanche probability per gate in non-Markov model is
\begin{equation}
p_n= F(p,p_a(j))+S(p,p_a(j))
\label{equ13}
\end{equation}

By keeping the first-order terms of $pa(j)$ only, we obtain the approximate equation of non-Markov model
\begin{equation}
p_n\cong p[1+\sum\limits_{j=1}^{n}(1-p)(p_a(j))].
\label{equ14}
\end{equation}

\begin{proof}
According to Equation \ref{equ13}, it is easy to get that $p_1=p[1+(1-p)p_a(1)]$, $p_2=p[1+(1-p)(p_a(1)+p_a(2))+(1-p)^2{p_a(1)}^2]\cong p[1+(1-p)(p_a(1)+p_a(2))]$, and $p_3=p[1+(1-p)(p_a(1)+p_a(2)+p_a(3))+(1-p)^2{p_a(1)}^2+2(1-p)^2p_a(1)p_a(2)+(1-p)^3{p_a(1)}^3] \cong p[1+(1-p)(p_a(1)+p_a(2)+p_a(3))]$. Thus, we assume that $p_{n-1}=p[1+(1-p)(p_a(1)+p_a(2))+\cdots+p_a(n-1)]$. If $p_n=p_{n-1}+p(1-p)p_a(n)$ holds, then Equation \ref{equ14} holds.

The right side of Equation \ref{equ13} includes two sets: Gate-0 avalanches in the first set $F(p,p_a(j))$ and does not avalanche in the second set $S(p,p_a(j))$.The second set can be expressed as $S(p,p_a(j))=(1-p)p_{n-1}$. The main difference of the first set from the second set is the afterpulsing effect ignited by Gate-0. If we separate out all terms due to the afterpulsing effect of Gate-0 (that is, terms with factors $p_a(1)$ of Gate-1, $p_a(2)$ of Gate-2, $\cdots$, $p_a(n)$ of Gate-$n$), the remaining terms of the first set can be expressed as $p\cdot p_{n-1}$. Terms with factor $p_a(1)$ of Gate-1 are $p(1-p)p_a(1)f_1(p,p_a(j))-p(1-p){f_1}^{'}(p,p_a(j))$, where $f_1(p,p_a(j))$ and ${f_1}^{'}(p,p_a(j))$ are the abbreviations for the remaining factors. According to the approximate convention, all terms with $p_a(j)$ in  $f_1(p,p_a(j))$ and ${f_1}^{'}(p,p_a(j))$ are abandoned. Then $f_1(p,p_a(j))$ and ${f_1}^{'}(p,p_a(j))$ approximate to the same expression 
\begin{displaymath}
f_1(p,p_a(j))\cong {f_1}^{'}(p,p_a(j))\cong {{n-1}\choose 0}p^{n-1}+{{n-1}\choose 1}p^{n-2}(1-p)+\cdots+{{n-1}\choose {n-1}}(1-p)^{n-1}=[p+(1-p)]^{n-1}=1,
\end{displaymath}
where ${n-1}\choose k$ is the binomial coefficient. Hence, $p(1-p)p_a(1)f_1(p,p_a(j))-p(1-p){f_1}^{'}(p,p_a(j))\cong 0$. Analogously, Terms with factor $p_a(2)$ of Gate-2, $p_a(3)$ of Gate-3, $\cdots$, $p_a(n-1)$ of Gate-$(n-1)$ all approximate to 0. As there is no term $-(1-p)p_a(n)$ for Gate-$n$, there remains one nonzero term $p(1-p)p_a(n)f_n(p,p_a(j))\cong p(1-p)p_a(n)$, where $f_n(p,p_a(j))\cong 1$ by the approximation convention. Thus,
\begin{displaymath}
\begin{aligned}
p_n&=F(p,p_a(j)+S(p,p_a(j))=p\cdot p_{n-1}+p(1-p)p_a(n)+(1-p)p_{n-1}=p_{n-1}+p(1-p)p_a(n)\\
&=p[1+\sum\limits_{j=1}^{n}(1-p)(p_a(j))].
\end{aligned}
\end{displaymath}
This completes the proof.

\end{proof}



\begin{thebibliography}{99}
\bibitem{Markovprocess}A process satisfies the Markov property if one can make predictions for the future of the process based solely on its present state, i. e., conditional on the present state of the system, its future and past are independent. Otherwise, if its future and past are correlated, the process is non-Markov. As shown in this letter, the afterpulsing probability distribution depends on the avalanching history. Thus afterpulsing effect satisfies non-Markov property.
\end{thebibliography}
\end{document}